\documentstyle[aasms4,12pt]{article}
\eqsecnum
\def\beq{\begin{equation}}
\def\eeq{\end{equation}}
\def\ref{\reference}
\def\simge{\mathrel{%
   \rlap{\raise 0.511ex \hbox{$>$}}{\lower 0.511ex \hbox{$\sim$}}}}
\def\simle{\mathrel{
   \rlap{\raise 0.511ex \hbox{$<$}}{\lower 0.511ex \hbox{$\sim$}}}}
\begin{document}

\title{OBSERVATIONS OF SPECTRAL AND TIME VARIABILITIES FROM MCG$-$2-58-22}

\author{CHUL-SUNG CHOI}
\affil{Korea Astronomy Observatory, 36-1 Hwaam, Yusong, Taejon 305-348,
Korea; cschoi@hanul.issa.re.kr}

\author{TADAYASU DOTANI}
\affil{Institute of Space and Astronautical Science, 3-1-1 Yoshinodai,
Sagamihara, Kanagawa 229-8510, Japan; dotani@astro.isas.ac.jp}

\author{INSU YI}
\affil{School of Physics, Korea Institute for Advanced Study, 207-43
Cheongryangri, Dongdaemun, Seoul 130-012, Korea; iyi@kias.re.kr}

\author{CHULHEE KIM}
\affil{Department of Earth Science Education, Chonbuk National University,
Chonju 516-756, Korea; chkim@astro.chonbuk.ac.kr}

%--------------------------------------------------------------------
\begin{abstract}

We present results from analysis of the X-ray archive data of 
MCG$-$2-58-22, acquired with ROSAT from 1991 to 1993 and with ASCA from
1993 to 1997.
By analyzing light curves, we find that MCG$-$2-58-22 shows a clear time
variability in X-ray flux.
The time scales of the variations range widely from $\sim\! 10^3$~s to more
than years.
Among the variations, a flare-like event overlaid on the gradual flux 
decrease from 1979 to 1993 is detected in the 1991 data; the flux has
increased at least by a factor of 3.
Combined analysis of the ROSAT and ASCA spectra shows that a simple
absorbed power-law does not fit the overall energy spectra, 
unless the column density lower than the Galactic value 
($3.5 \times 10^{20}$ cm$^{-2}$) is adopted.
We find the clear time variability of the spectra in the energy range of 
0.1--2.0 keV\@.
The spectral shape with respect to an adopted model continuum is generally
correlated with their flux level.
However, the flux variation does not result in any significant influences on 
their spectra in the energy range of 2--10 keV\@.
As reported previously, there is an iron line centered at $\sim 6.3$ keV
(which is not corrected for the red-shift effect),
but we find that the line width is broad $\sigma = 0.9^{+0.6}_{-0.3}$ keV
with a single Gaussian model.
The implications of these observational results are discussed in terms of
a supermassive black hole model and accretion flow dynamics near the central
black hole.

\end{abstract}


%--------------------------------------------------------------------
\section{INTRODUCTION}

MCG$-$2-58-22 (Mrk 926) is an active galactic nucleus (AGN) optically 
classified as Seyfert 1 or 1.2 galaxy (Whittle 1992; Kotilainen \& Martin 
1994).
It is located at a distance of 284 Mpc (z = 0.04732
and H$_0$ = 50 km s$^{-1}$ Mpc$^{-1}$; Huchra et al.\ 1993).
In X-ray band, the source has been observed frequently for more than
two decades from the Einstein observatory to ASCA\@.
Many observational results, however, still remain either to be understood 
or to be confirmed.
Among those, the first is the ``soft X-ray excess" problem, 
which in general is defined as emission that exceeds the extrapolation 
from the observed hard X-ray power-law continuum.
Some authors have argued that they found an excess emission below
1$-$2 keV in their spectra of MCG$-$2-58-22 (e.g., Turner et al.\ 1991; 
Ghosh \& Soundararajaperumal 1992; Piro, Matt, \& Ricci 1997), but different 
authors have disagreed (Mushotzky 1984; Turner \& Pounds 1989; Turner, George,
\& Mushotzky 1993; Reynolds 1997).
In addition, according to Nandra \& Pounds (1994), this source exhibits
a ``hard tail" (or hard X-ray excess) above 10 keV\@.
This soft and/or hard excess is model dependent and is especially sensitive to 
a continuum slope.  
Hence, an accurate measurement of continuum slope in a wide energy
range is essential for understanding these excess phenomena.
At present, it is not yet clear whether the soft excess in MCG$-$2-58-22
is transient or stable or whether the excess itself exists.

The second problem is that the measured continuum slope has so far been 
different from observatory to observatory.
For instance, it has been reported that the time-averaged continuum of
MCG$-$2-58-22 can be described by a simple power-law with a spectral
index of $\Gamma \approx 2.1$ in a ROSAT energy band  and with 
$\Gamma \approx 1.75$ in an ASCA energy range (e.g., Turner, George, \&  
Mushotzky 1993; Weaver et al.\ 1995; Piro, Matt, \& Ricci 1997).
The energy-dependent spectral slope is unclear and it is yet to be seen
whether it is real or due to a limited pass-band.
There is an iron line centered at $\sim\! 6.2$ keV (e.g., Nandra \& Pounds
1994 and Weaver et al.\ 1995).
Moreover, according to Turner et al.\ (1991) and Turner, George, \& 
Mushotzky (1993), there is a soft X-ray line at $\sim\! 0.8$ keV\@.
Finally, X-ray flux of MCG$-$2-58-22 is known to vary by a factor of 4,
from $\sim\! 1 \times 10^{44}$ to $\sim\! 4 \times 10^{44}$ ergs s$^{-1}$ 
(2$-$10 keV), but little is known about the nature of the time
variations.
Ghosh \& Soundararajaperumal (1992) examined the spectral variability with 
EXOSAT data, and they concluded that soft X-ray is subject to change with
the flux variation, and consequently magnitude of the soft excess
also varies.
This result is still remained to be confirmed.

In this paper, we examine these problems in detail.
In \S\ 2 we describe the ROSAT and ASCA observations as well as the data
reduction procedures.
In \S\ 3 we analyze light curves and obtain some new results for their time 
variability. 
Fitting of the combined ROSAT and ASCA spectra is performed in \S\ 4.
Using the model of the combined spectra, we also investigate a spectral
variability in \S\ 4 for data that were obtained from different observations.
Finally, we discuss the results of our study in \S\ 5.

%--------------------------------------------------------------------
\section{OBSERVATIONS AND DATA REDUCTION}

ROSAT observations of MCG$-$2-58-22 were made four times from 1991
to 1993 as listed in Table 1.
These pointed observations were carried out with the Position Sensitive 
Proportional Counter (PSPC-B) with an on-axis spatial resolution of 
about 20 arcsec (Pfefferman et al.\ 1986).
PSPC is sensitive in a 0.1$-$2.0 keV energy range and has a field of 
view of 2 degrees.
Energy resolution of the PSPC is approximated by ${\Delta E}/E =
0.43(E/0.93)^{-0.5}$ in a full-width at half-maximum (FWHM), where 
$E$ is in keV\@.

The ROSAT data were retrieved through the High Energy Astrophysics
Science Archive Research Center (HEASARC) on-line service, provided by
the NASA/Goddard Space Flight Center.
Among the available data, both data sets of 1991 November 21 (rp700107) 
and 1993 May 21 (rp701250) were partly analyzed and published by, e.g.,
Turner, George, \& Mushotzky (1993) and Piro, Matt, \& Ricci (1997), 
but other data have not been yet. 
All the available data are included in this study.
Data reduction was performed with the FTOOLS v4.2 software package
in our study (Turner 1996).
Source photons are extracted from a circular region of about 
2 arcmin centered on the source, and the background data are obtained  
from annular source-free regions in the same field of view.

MCG$-$2-58-22 was also observed by ASCA three times during the period of 
1993 to 1997 (Table 1).
Particularly, ASCA and ROSAT were pointed at this source simultaneously on
1993 May 25$-$26 for about 2.3 hrs.
Some results of this ASCA data set were already published by several
researchers, including Weaver et al.\ (1995), 
Reynolds (1997) and Nandra et al.\ (1997a, 1997b).
Nevertheless, we use this data set in order to do an analysis for combined
ROSAT and ASCA spectra.
However, other sets of ASCA data still remain to be analyzed.  

The ASCA satellite has 4 focal plane instruments, two of which are
Solid-State Imaging Spectrometers (SIS0 \& SIS1) and the other two
are Gas Imaging Spectrometers (GIS2 \& GIS3).
Each of the SISs consists of 4 CCD chips with an FWHM energy resolution
of $\sim\! 60-120$ eV in the 0.4$-$10 keV energy range at launch time
(Burke et al.\ 1993), while the GISs have an energy resolution of 
$\sim\! 200-600$ eV in the 0.8$-$10 keV range (Ohashi et al.\ 1996; 
Makishima et al.\ 1996). 
The SISs in 1 CCD mode provide an 11 arcmin $\times$ 11 arcmin square
field of view, while the GISs give a circular field of view with a
diameter of 50 arcmin, regardless of their observational mode.

We acquired ASCA data from the HEASARC public archives. 
In order to avoid possible X-ray contamination from the bright Earth and 
regions of high particle background, we apply standard data-screening
procedures: 
(1) data are rejected for SISs and GISs when the pointing
direction of the telescope is less than 30$\arcdeg$ and 8$\arcdeg$ from
the Earth's limb, respectively, 
(2) only the regions where the radiation-belt monitor has a count rate 
less than 300 count/s are selected for SISs,
(3) regions of cutoff rigidity larger than 10 GeV/c are selected for
both of the instruments. 
Hot and flickering pixels are also removed from the SIS data.
Then, source photons are extracted from a circular region of about
4 arcmin centered on the source. 
Background data are extracted from the high-latitude, blank sky data
publicly available.

%--------------------------------------------------------------------
\section{LIGHT CURVES AND TIME VARIABILITIES}

It is well known that time variability in X-ray intensity is a common 
characteristic of AGNs, which ranges from hours to years.
According to the recent studies by Nandra et al.\ (1997a) and Ptak et al.\ 
(1998), there is a clear anti-correlation between X-ray luminosities and 
magnitude of time variabilities in Seyfert 1 galaxies.
MCG$-$2-58-22 is a bright Seyfert ($L_X \sim 10^{44}$ ergs s$^{-1}$)
deficient of rapid time variations, although it is known to be
variable by a factor of 4 on time scales of years.
We study the variability of MCG$-$2-58-22 on both short and long
time scales.

Figure 1 shows six light curves extracted from the ROSAT/PSPC
(left panels; 0.1$-$2.0 keV) and the ASCA/GIS data
(right panels; 0.8$-$10 keV). 
In the figure, GIS2 and GIS3 data were averaged and each data point was
binned in 300 s intervals.
The X-ray background, which is negligible in the plot, was not subtracted. 
Among the light curves, we clearly see a steady increase of the X-ray 
flux (0.1$-$2.0 keV) in the ROSAT light curve of 1993 May 21$-$26
(MJD 49128$-$49138; left-middle panel).
We performed the $\chi^2$ test against the hypothesis of a constant
X-ray flux, which resulted in a $\chi^2$ value of $\chi^2/\nu$
(degree of freedom) = 325.2/110.
The hypothesis is clearly rejected; X-ray flux from MCG$-$2-58-22
is variable within the time scale of a few days.
To understand the nature of the variability, we investigated 
energy-dependent light curves obtained in the 0.1$-$0.9 keV band and
the 0.9$-$2.0 keV band.
We confirmed that a similar pattern of the flux increase was seen in the 
both energy bands.
The calculated $\chi^2$ values against a constant flux were 
$\chi^2/\nu$ = 213.6/110 (0.1$-$0.9 keV) and $\chi^2/\nu$ = 230.6/110
(0.9$-$2.0 keV), respectively.

Exposure times of the ROSAT observations of 1991 November 21 
(MJD 48581) and 1993 December 1 (MJD 49322) were too short for
a $\chi^2$ test.
Instead, we closely looked at the light curves.
From the visual inspection, we found a significant short-term variation
in the light curve of 1991 November 21 (Figure 2).
It is clear from the figure that there is a sudden increase and decrease of 
flux, from 3.9 counts/s to 4.4 counts/s, at the observation time of 19.2 hr.
This variation has a time scale of $\sim\!10^3$~s.
We also closely examined the light curve of 1993 December 1, but could
not find such a rapid variation.
We calculated the $\chi^2$ values (against the constant hypothesis)
for the three sets of the ASCA light curve in Figure 1.
The calculated $\chi^2$ values for the ASCA light curves are
$\chi^2/\nu$ = 103.9/87 for the 1993 May observation (MJD 49133--49134), 
and 149.6/98 and 106.6/102, respectively, for the 1997 June and December 
observations.
Only the light curve in 1997 June shows statistically significant
time variation.
The variation may have a similar time scale as the ROSAT data in
1993 May, although statistics of the ASCA data are not good enough to 
investigate the energy dependence.
We also see a factor of 2 change in the GIS flux (0.8$-$10 keV) 
between 1993 and 1997 observations.
ASCA light curve of 1993 May was also analyzed by Nandra et al.\ (1997a),
and they obtained the same conclusion as ours.

In addition to the above light curves, we have also made a long-term light
curve (2$-$10 keV)\@.
As explained in the subsequent section, the spectral shape of MCG$-$2-58-22 
is weakly dependent on its flux level in a 2$-$ 10 keV band.
Therefore, we assumed a power-law ($\Gamma = 1.7$) modified by the
low-energy absorption (N$_H = 2.3 \times 10^{20}$ cm$^{-2}$),
which explains both ASCA and ROSAT energy spectra simultaneously 
(see Table 2 and \S\ 4.3 to convert the ROSAT count rate into a
2$-$10 keV flux).
For the long-term light curve, we have also included ROSAT all-sky survey 
data (Voges et al.\ 1999), as well as those calculated from previous 
studies (open circles in the figure; Turner et al.\ 1991; Ghosh \&
Soundararajaperumal 1992; Nandra \& Pounds 1994).
The result is shown in Figure 3, which covers about 18 yrs from 1979 to 1997.
We see a long-term change of the flux, which decreased gradually from 1979 
through 1993 but increased afterward.
One interesting point is that there is a flare-like event in 1991 November;
the flux is at least three times larger than that expected from the 
long-term trend between 1979 and 1993.
Because we have only a single set of observation, it is difficult to 
estimate the true magnitude and duration of the flare-like event.
However, the time scale may be less than a few years and the flux
increased by at least a factor of 3.

%--------------------------------------------------------------------
\section{SPECTRAL ANALYSES AND THE RESULTS}
\subsection{Fitting of Combined ROSAT and ASCA Spectra}

It has been reported that the time-averaged continuum of MCG$-$2-58-22
can be described by a simple power-law model, i.e., power-law $\times$ 
Galactic absorption, with a spectral index of $\Gamma \approx 2.1$ 
in the ROSAT energy band and of $\Gamma \approx 1.75$ in the ASCA energy
band (e.g., Turner, George, \& Mushotzky 1993 and Weaver et al.\ 1995).
In addition, there is an iron line feature around 6$-$7 keV (Nandra \&
Pounds 1994; Weaver et al.\ 1995).
In order to examine whether this energy-dependent spectral slope is real, 
we combine ROSAT and  ASCA spectra.
As mentioned in \S\ 2, the two observatories pointed at MCG$-$2-58-22
at the same time for about 2.3 hrs.
Therefore, we use these data to avoid any temporal spectral change effect.
The spectra shown in Figure 4 were extracted separately from the PSPC, SIS 
and GIS data, and the background was subtracted.
In this process, we added the two energy spectra from SIS0 and SIS1, and
simultaneously from GIS2 and GIS3, to improve statistics.

We first attempt to fit the combined spectra with the simple power-law
model (model 1).
In this fit, normalization of each instrument is allowed to vary
independently, but other parameters such as spectral index $\Gamma$
and hydrogen equivalent column density N$_H$ are linked together.
The resulting $\chi^2$ in Table 2 implies that the combined spectra can be
fitted by a single power-law in a wide range of 0.1$-$10 keV relatively well.
Some residual structure is recognized around 6$-$7 keV, but the spectral 
ratio between the data and the model continuum does not show any 
significant soft X-ray excess (upper panel of Figure 5).
This means that there may be little difference in spectral slope between the 
soft (ROSAT) and the hard (ASCA) spectra.
However, the measured column density in the line of sight to the source, 
N$_H = 2.2^{+0.4}_{-0.5} \times 10^{20}$ cm$^{-2}$, is slightly smaller 
than the Galactic value $3.5 \times 10^{20}$ cm$^{-2}$ 
(e.g., Piro, Matt, \& Ricci 1997).
If N$_H$ is fixed to the Galactic value (model 2), it results in 
a soft X-ray excess below about 0.5 keV as shown 
in the lower panel of Figure 5.
Increase of $\chi^2$ was also significant.
This result suggests that if one sticks to the Galactic absorption column, 
one has to include an additional spectral component to explain 
the soft excess.
As an alternative test, we applied different absorption model such as an 
ionized or warm absorber model (``absori" in the XANADU/XSPEC package),
but found it does not affect the above results significantly.
For instance, when the parameters of ``absori" were allowed to be free, 
spectral index and column density parameters converged to the values 
of model 1, which uses photo-electric absorption
cross-sections of Morrison \& McCammon (1983).

Although the simple power-law model gave an acceptable fit from
a statistical point of view, we note a residual structure around
6$-$7 keV, which is broad and weak but significant (Figure 5).
This is suggestive of an emission line.
Because the presence of an emission line might affect the
estimation of the power-law slope and the hydrogen equivalent column
density, we tried to fit a third model to the combined data.
Considering both this structure and previous reports on the detection of 
iron line from MCG$-$2-58-22, we add a Gaussian to the power-law model:
(power-law + Gaussian) $\times$ low-energy absorption (model 3).
We tried only the simplest line model because of relatively poor 
statistics of the data.
Solid lines in Figure 4 represent the best-fit model, and the best-fit
parameter values and their uncertainties at the 90\% confidence limit 
are listed in Table 2.
In this fit, we did not correct the red-shift effect and assumed a broad 
iron line and fixed the Gaussian width to $\sigma = 0.9$ keV (\S\ 4.2).
By including the Gaussian, we could obtain a substantially improved fit as
shown in Table 2.
Based on the line parameters, we calculate the line flux to be
$1.1^{+0.4}_{-0.5} \times 10^{-12}$ ergs cm$^{-2}$ s$^{-1}$.
The best-fit power-law slope is similar to that of model 2, and the hydrogen
equivalent column density is found to be 
N$_H = 2.3^{+0.5}_{-0.3} \times 10^{20}$ cm$^{-2}$, which is 
slightly smaller than the Galactic value.
Thus the inclusion of an iron-K line does not resolve the discrepancy
between the best-fitting column density and the Galactic column.
We will return to this problem in \S\ 4.3.

%--------------------------------------------------------------------
\subsection{Iron Line Study}

According to the recent study by Nandra et al.\ (1997b), most of the Seyfert 1
galaxies observed with ASCA show broad iron K${\alpha}$ lines with 
a mean width of $\sigma = 0.43 \pm 0.12$ keV\@.
In addition, some UV emission lines, Ly${\alpha}$ and C$_{\rm IV}$ (1549 \AA),
are measured to be broad for MCG$-$2-58-22 ($\sim\! 5000$ km s$^{-1}$ at
FWHM in the rest frame; T\"urler \& Courvoisier 1998).
In practice, an accurate determination of the line parameters is difficult
with the above ASCA spectrum, because of the weakness of the line structure 
and the relatively poor data statistics in higher energy range. 
Here we examine iron line properties further with ASCA/SIS data that obtained 
in 1997 June and 1997 December for longer exposure.
Because we found no significant difference between the energy spectra
at these two epochs, we added the two spectra to improve the statistics.

In the course of data analysis, we realized that the radiation damage
effect is significant in the lower energy range ($<\!1.5$ keV) 
of SIS data (Hwang et al.\ 1999; Yaqoob 2000).
To demonstrate the effect, we first fit the model 3 (Table 2)
to the GIS data with all the parameters fixed except for the normalization
and confirmed that the model can fit the data very well.
Then, we calculated a PHA ratio of the SIS spectrum to model 3, 
which is shown in Figure 6.
As seen from the figure, SIS data become significantly smaller than
the model below 1.5 keV\@.
This is due to the radiation damage to SIS\@.
Because this effect is not included in the current calibration 
data yet, we just discard the SIS data below 1.5 keV\@.

To see the iron line structure first, we calculate spectral ratio between
the summed SIS data and the model continuum, which is shown in Figure 7.
This figure shows a broad structure between 5--8 keV,
implying the presence of a broad iron line.
If we apply a single Gaussian model to the line, we obtain an 
acceptable fit with the best-fit line width of 
$\sigma = 0.9^{+0.6}_{-0.3}$ keV (Table 3).
On the other hand, if we apply a disk line model to investigate the
origin of the broadness, we get a similar goodness of fit (Table 3).
We have fixed some of the parameters in the disk line model, to which 
the $\chi^2$ value is not very sensitive, in the fitting.
The fitting results indicate that, although the line profile is very 
different between the broad Gaussian and the disk line models, current data 
do not have good enough statistics to distinguish the line profiles.
In fact, we could not constrain the disk line parameters very well.
Because there already exists detailed study of the iron feature
in MCG$-$2-58-22 (Weaver et al.\ 1995; Nandra et al.\ 1997b), 
we do not further elaborate on the study of the iron feature in the 
present paper.

%--------------------------------------------------------------------
\subsection{Soft-Band Variability} 

As we can see in Figure 7, there is little difference in the continuum
spectrum between the 1993 and the 1997 data.
This means that although there is a factor of 2 change in X-ray flux,
this flux variation does not accompany any spectral change in 2$-$10 keV\@.
However, this does not necessarily mean that the spectral shape is
also stable in lower energy band.
Because ASCA/SIS cannot be used to look at the long-term spectral change
below 2 keV due to the accumulating radiation damage, we concentrate on 
the ROSAT data in this subsection.

As we did in \S\ 4.2, we calculate spectral ratios for the ROSAT data
taking model 3 (with the best-fit parameters to the combined ROSAT
and ASCA data) as a reference (Figure 8).
The ratio may be constant for 1993 December data, but a clear structure 
is seen in 1991 November data. A similar structure, but less pronounced,
is also recognized in 1993 May data.
It is clear from Figure 8 that either (or both) of the column density
or the power-law index is time variable.
To identify the variable component, we fit a power-law model with the
low-energy absorption to the ROSAT PSPC spectra, and calculated
confidence contour plots.
The results are summarized in Table 4 and in Figure 9.
For clarity, we did not plot the confidence contour for the data in
1993 December, which gave only a very loose constraint to the parameters.

We can see from the confidence contours that the energy spectra
clearly have a steeper slope than that determined by the ASCA data.
The photon index determined by ASCA in 1.5$-$10 keV is $1.73\pm0.04$, 
whereas the PSPC required the photon index to be larger than 1.8 
(1993 May 24-25) and 1.9 (1993 May 21-23, 1991 November 21).
Most probable photon index in PSPC band is 2.1.
As explained in \S\ 4.1, we succeeded in fitting the combined ASCA
and ROSAT spectra by a simple absorbed power-law model.
The result is not necessarily in contradiction to the result in this
subsection, because the statistics of the combined ASCA/ROSAT spectra
were not very good.
Furthermore, even if the power-law slope in the ROSAT band is allowed to
be time variable, acceptable range of the absorption column is, 
although marginal, smaller than the Galactic value at least for the data 
in 1991 November.
We interpret this result as implying that the energy spectrum of 
MCG$-$2-58-22 in ROSAT band is not a simple absorbed power-law, but has 
more complex structures.
The structures may be time variable, and are probably correlated with 
the flux level. 
When the X-ray flux is large, the soft band spectrum tends to
become steeper, which results from either a larger photon index or
a lower column density.

%--------------------------------------------------------------------
\section{DISCUSSION}

We have analyzed the ROSAT and ASCA archive data of MCG$-$2-58-22.
From the long-term light curve covering from 1979 through 1993, 
we found a flare event in 1991 November overlaid on the
gradual change of the X-ray flux.
The X-ray flux increased at least a factor of 3 during the flare.
Although the time scale of the flare is not known, it is less
than about a year.
We also detected a very rapid time variation of a relative amplitude
of about 10\% with a time scale of $10^3$~s.
Energy spectra of MCG$-$2-58-22 were found to be rather complex.
Iron-K line is significantly broad, and soft band spectra (0.1$-$2 keV)
contain some structures, which cannot be reproduced by a simple 
absorbed power-law.
But the energy resolution and statistics of the current data may not be
enough to study the iron line and the low energy structures in detail.
We discuss the origin of the time variabilities first and then 
the structures in the energy spectra.

\subsection{Origins of Time Variabilities in MCG$-$2-58-22}

The long term modulations with occasional flares in the light curves of 
some AGNs are mostly seen in the optical range.   The long term X-ray
light curve of MCG$-$2-58-22 we obtained appears interestingly
similar to some of the long term optical light curves. 
The large amplitude flares and gradual long term modulations could 
imply that these systems are undergoing at least two distinctly different
physical processes differing in time scales and spatial locations.
Webb (1990) reported the results of 61 years of optical observations of
3C120 and claimed that a gradual and possibly sinusoidal variability 
component of period 12.43 yrs as well as high amplitude flares on much 
shorter time scales. 
This behavior is very similar to that of MCG$-$2-58-22 we obtained.
It is also interesting to point out that Dibai \& Lyutyi (1984) claimed that
AGN such as NGC 3516 and others essentially have two variability time scales,
10$-$30 day flares and 10$-$30 yr sinusoidal variation, which appear to be 
similar to those in other long-term variability studies.

If the variabilities arise from the viscous and dynamical processes in
accretion disks in galactic nuclei, the characteristic time scales can be as
long as years which correspond to the outermost regions of accretion disks.
Observations in various bands suggest that a wide range of variability time
scales exist, which in turn indicates a wide range of distance scales
associated with various emission components. It has been often discussed that
the long term ($\sim$ yrs $-$ a few $\times$ 10 yrs) variabilities are
caused by some types of tidal interaction while the short term variabilities
have been attributed to the accretion emission regions.

One of the possible origins of the long term variabilities is the dynamic
changes in emissions regions in terms of the re-distribution of
matter and resulting variable emission and/or self-absorption.
There are a number of relevant time scales operating in the accretion disk
around a massive black hole in the galactic nucleus
(Frank et al.\ 1992 and references therein).
The typical dynamical time scale for a Keplerian disk is
\begin{equation}
t_{dyn}\sim 3\times 10^{-2}(M/10^7\;M_{\odot})^{-1/2}(R/10^{15}\; 
{\rm cm})^{3/2}~{\rm yr},
\end{equation}
where $M$ is the mass of the black hole and $R$ is the distance from  
the black hole. 
This time scale would provide a long term variability of a few yr to
ten yrs for $R>10^{16}$ cm, although the exact mechanisms responsible for such
variabilities remain somewhat uncertain. The viscous time scale on which the
disk material's density is modulated is much longer than the dynamical time
scale, i.e.,
\begin{eqnarray*}
t_{vis} & \sim & R^2/\nu \\
        & \sim & \alpha^{-1}(H/R)^{-2}t_{dyn} \\
        & \sim & 10^3 (\alpha/0.1)^{-1}\left[(H/R)/0.1\right]^{-2} t_{dyn} \\
        & \gg  & t_{dyn},
\end{eqnarray*}
where $\nu$ is the kinematic viscosity coefficient, $\alpha\sim 0.1-1$ is the
dimensionless viscosity parameter, and $H$ is the disk's vertical pressure
scale height.
The thermal fluctuations in the disk operate on the thermal time scale
\begin{equation}
t_{th}\sim 10 (\alpha/0.1)^{-1}t_{dyn},
\end{equation}
which falls between the viscous time scale and the dynamical time scale.
If the disk experiences a viscous-thermal instability, which is caused by the
sudden change of hydrogen ionization in the disk at an effective disk
temperature $T_{eff}\sim 4000$ K\@. 
The viscous-thermal instability time scale is roughly estimated as
\begin{equation}
t_{vis-th}\sim 3\times 10^2(\alpha/0.1)^{-1}(M/10^7M_{\odot})^{1/3}
({\dot M}/10^{-4}M_{\odot}/{\rm yr})^{1/3},
\end{equation}
which is a relevant time scale for mass accretion rate modulation on long time
scales.
These time scales imply that there in principle exists a large range of time
scales on which mass accretion rate (${\dot M}$) and emission can be modulated.

The origin of flare-like modulations such as the one we detected in 
MCG$-$2-58-22 have not been convincingly accounted for.
One of the possible mechanisms is the disruption of stars in the
tidal field of the massive black hole (e.g., Cannizzo et al.\ 1990).
In a closely related process, a star could plunge into the accretion disk and
drag-heat the accretion disk creating a sudden episode of heating and emission
(e.g., Hall et al.\ 1996 and references therein),
although the frequency, duration, and intensity of such encounters depend on
many detailed dynamical as well as plasma physical details which are beyond the
scope of this paper.  As in some Galactic binary systems, the accretion disk
could have a hot scattering corona which Compton-upscatters the lower energy
photons emitted by the accretion disks.  Any such coronal model for X-ray
emission requires an electron acceleration mechanism such as magnetic flares
and dissipation driven by magnetic field buoyancy and reconnection.
Such a process is inherently flare-like and it occurs on relatively short time
scales (e.g., Di Matteo, Blackman, \& Fabian 1997).
The detailed theoretical discussions and their observational signatures
will be presented elsewhere.

We detected very rapid time variation as short as $10^3$ s from MCG$-$2-58-22.
Nandra et al.\ (1997a) studied such ultra-rapid variability 
for a sample of 18 Seyfert 1 galaxies using the ASCA data.
They detected significant power at frequencies greater than $10^{-3}$ Hz
in at least five sources.
However, unequal sampling and poor statistics of the data make it 
difficult to detect such ultra-rapid variabilities in most cases.
In fact, although their analysis sample included MCG$-$2-58-22, 
they could not detect ultra-rapid variability.
It is known that there is a good correlation between the X-ray
variability amplitude and the source luminosity (Nandra et al.\ 1997a;
Turner et al.\ 1999).
Because MCG$-$2-58-22 is a bright Seyfert ($L_{\rm 2-10keV} = 
1-4 \times 10^{44}$ ergs s$^{-1}$), it shows relatively low amplitude of 
time variation.
However, we found an ultra-rapid variability from MCG$-$2-58-22.
This may be interpreted as implying that high frequency variabilities 
($>10^{-3}$ Hz) is common to Seyfert galaxies, but the present
observations do not have a sensitivity high enough to detect them.
Most rapid time variations in X-ray flux may be expected from the inner 
most region of the accretion disk, which corresponds to 3 times
the Schwarzschild radius ($6GM/c^2$).
The dynamical time scale ($t_{dyn}^{in}$) at this radius is:
\begin{equation}
t_{dyn}^{in}\sim 8\times 10^2 (M/10^7\;M_{\odot})~{\rm s}.
\end{equation}
This is comparable to the observed time scale.
Although we do not know the mass of the central black hole in
MCG$-$2-58-22, the ultra-rapid variabilities ($\sim\!10^3$ s)
may be produced at the vicinity of the event horizon of the black hole.

\subsection{Structures in the Energy Spectra}

We found that the energy spectra of MCG$-$2-58-22 in 0.1$-$10 keV cannot be
reproduced all the times by a simple power-law modified by the low
energy absorption.
Energy spectra above $\sim$2 keV gives a photon index of 1.73, whereas
lower energy spectra (0.1$-$2 keV) prefer a photon index of 2.1.
Furthermore, if we fit a power-law ($\Gamma = 2.1$) to 0.1$-$2 keV spectra,
the best-fitting absorption column becomes lower than the Galactic
value.
If we force the Galactic absorption, some of the energy spectra cannot be
fit by the absorbed power-law model.
We also found that the low energy spectra of MCG$-$2-58-22 is not
stable, but show significant time variations.
Therefore, depending on the statistics of the data, an simple
power-law with a low energy absorption may give statistically acceptable
fit to 0.1$-$10 keV data.
However, if the data have good statistics, such a simple model may be
rejected from the statistical point of view.
We consider a possibility that the time variations of the low energy
structures may be the main source of complexities in the spectral shape of
MCG$-$2-58-22.

There are several possible origins for the low energy structures.
Low energy spectra of selected Seyfert 1 galaxies (including MCG$-$2-58-22)
are studied in reference to the Ginga analysis (Piro et al.\ 1997).
The models they tried include ``warm absorber'', ``reflection from
mildly ionized material'', and a blackbody components, but
none of these models gave satisfactory fits to all the data;
combination of two components might be needed.
Zdziarski et al.\ (1999) found a very strong correlation between 
the intrinsic spectral slopes and the amount of Compton reflection
from a cold medium in Seyferts and in the hard state of X-ray
binaries.
Objects with softer intrinsic spectra tend to have much stronger
reflection structure.
Because MCG$-$2-58-22 has a relatively hard intrinsic spectra 
($\Gamma = 1.67\pm0.08$; Nandra \& Pounds 1994), only a small reflection
structure may be expected in its energy spectrum.
In fact, Nandra \& Pounds (1994) found little reflection structure
in the Ginga spectrum of MCG$-$2-58-22.
Although these considerations may prefer such an interpretation that 
the structures in the energy spectra of MCG$-$2-58-22 are more closely related 
to the warm absorber, we consider that the available data sets are
not enough to draw clear conclusion.

We detected a broad iron line from MCG$-$2-58-22.
Both a simple Gaussian and a disk line model gave a similarly good fit to
the data, and we could not investigate the line profile in detail.
However, as shown in Figure 7, the line is clearly dominated 
by a broad component, and the contribution of a narrow line, if present, 
is small.
Iron-K line profiles were investigated as a function of
X-ray luminosity in AGNs (Nandra et al.\ 1997c).
The line strength clearly decreases with increasing luminosity.
Furthermore, the line profiles also changed; narrow Gaussian component
largely disappeared above $L_X \sim 10^{45}$ ergs s$^{-1}$ (2--10 keV)\@.
MCG$-$2-58-22 has an X-ray luminosity of $1-4 \times 10^{44}$ ergs s$^{-1}$.
Thus the line profile may lack the narrow component, which is
consistent to the line profile we obtained (Figure 7).

%--------------------------------------------------------------------
\section{SUMMARY AND CONCLUSION}
Our analyses of MCG$-$2-58-22 X-ray data have yielded the following results:
\begin{enumerate}
\item The 1979--1997 X-ray light curve clearly exhibits a long term 
      variability.   A flare-like event is detected in 1991 overlaid over
      the gradual flux variations.
\item The observed variability time scales span a wide range from $10^3$~s 
      to years.   The most rapid variability we detected is as short as 
      $10^3$~s, which could be related to the dynamical time scale at the 
      inner most region of the accretion disk.
\item The combined ROSAT and ASCA energy spectra may not be reproduced by an
      absorbed power-law model, so long as the Galactic absorption of 
      $3.5 \times 10^{20}$ cm$^{-2}$ is adopted.   Energy spectra in ROSAT band
      (0.1--2.0 keV) requires either a steeper continuum or a lower absorption.
\item The observed spectra in the energy range of 0.1--2.0 keV are clearly time
      variable.
\item There exists a broad iron line centered at $\sim 6.3$ keV.
\end{enumerate}

Origin of the short term flare event on top of the gradual X-ray flux 
modulation remains unresolved.
It is interesting that similar flare-like events have been reported in the
optical light curves of some AGNs which also show gradual modulations on time
scales of years and tens of years.
There exist some interesting physical time scales which naturally arise in
accretion disks around supermassive black holes.
However, it has not been accounted for how the flares of these time scales
are produced.

Energy spectra of MCG$-$2-58-22 are quite complex.
Iron-K line is noticeably broad, and soft band spectra (0.1$-$2 keV)
contain some significant structures, which cannot be accounted for
by a simple absorbed power-law.
Although it is possible that the low energy structures are more closely
related to the warm absorber, the presently available data sets are
not good enough to draw a definite conclusion.

%--------------------------------------------------------------------
\acknowledgments

%--------------------------------------------------------------------

%--------------------------------------------------------------------
\clearpage

\begin{table}
\begin{center}
\caption{THE ROSAT AND ASCA OBSERVATIONS OF MCG--2-58-22}
\vspace*{0.7cm}

\begin{tabular}{llccc}  \tableline \tableline
\multicolumn{1}{l}{DATA$^1$} &
\multicolumn{1}{l}{OBSERVATION} &
\multicolumn{1}{c}{INSTRUMENT} &
\multicolumn{1}{c}{EXPOSURE$^2$} &
\multicolumn{1}{c}{COUNT RATE$^3$} \\ 
  & (DATE) &  & (ks) & (c/s) \\ \tableline 

rp700107 & 1991 November 21 &  PSPC-B & 3.5 & 4.0 \\
rp701250 & 1993 May 21--25 & PSPC-B & 18.7 & 0.9 \\
rp700998 & 1993 May 24--26 & PSPC-B & 10.6 & 1.0 \\
rp701364 & 1993 December 1 & PSPC-B & 2.2 & 0.7 \\
ad70004000 & 1993 May 25--26 & GIS, SIS & 23.3 & 0.4 \\
ad75049000 & 1997 June 1--2   & GIS, SIS & 26.3 & 0.9 \\
ad75049010 & 1997 December 15--17 & GIS, SIS & 29.8 & 0.9 \\ 
\tableline

\multicolumn{5}{l}{$^1$ Sequential number of archival data.}\\
\multicolumn{5}{l}{$^2$ Net exposure times after the data screening. In the
case of ASCA observations, the exposures}\\
\multicolumn{5}{l}{are from the GIS.}\\
\multicolumn{5}{l}{$^3$ Mean source count rates before the subtraction of the 
background are from PSPC and GIS.}\\

\end{tabular}
\end{center}
\end{table}

%--------------------------------------------------------------------
\clearpage

\begin{table}
\begin{center}
\caption{BEST-FIT RESULTS FOR THE COMBINED ROSAT AND ASCA SPECTRA}
\vspace*{0.7cm}

\begin{tabular}{lccc} \tableline \tableline
PARAMETERS &  & BEST-FIT VALUES & \\ \cline{2-4}
           & model 1$^a$ &  model 2$^b$ & model 3$^c$ \\ 

\tableline

$\Gamma$   & 1.69$^{+0.03}_{-0.03}$   & 1.73$^{+0.02}_{-0.02}$ 
           & 1.73$^{+0.04}_{-0.04}$ \\
N$_H$ (10$^{20}$ cm$^{-2}$)        & 2.2$^{+0.4}_{-0.5}$ & 3.51 
                                   & 2.3$^{+0.5}_{-0.3}$ \\
Normalization for PSPC (10$^{-3}$) & 4.0$^{+0.2}_{-0.2}$  
      &  4.2$^{+0.2}_{-0.2}$    & 4.0$^{+0.2}_{-0.2}$ \\
Norm. for SIS (10$^{-3}$)          & 3.73$^{+0.07}_{-0.07}$  
      &  3.86$^{+0.07}_{-0.07}$    & 3.77$^{+0.08}_{-0.08}$ \\
Norm. for GIS (10$^{-3}$)          & 4.6$^{+0.2}_{-0.1}$  
      &  4.8$^{+0.1}_{-0.1}$    & 4.7$^{+0.2}_{-0.1}$ \\
E$_K$$^d$ (keV)  & \nodata & \nodata & 6.0$^{+0.6}_{-0.5}$ \\
$\sigma$$^e$     & \nodata & \nodata & 0.9 \\
I$_{Fe}^f$ ($10^{-4}$ photons cm$^{-2}$ s$^{-1}$) & \nodata &
        \nodata & 1.1$^{+0.4}_{-0.3}$ \\
EW$^g$ (keV)     & \nodata & \nodata & 0.6$^{+0.2}_{-0.3}$ \\
$\chi^2/\nu$                       & 82.4/114 & 106.1/115 
                                   & 69.1/112 \\
Flux$^h$ (0.1$-$2.0 keV) & 1.1 & \nodata 
                                              & 1.1 \\
Flux$^h$ (2.0$-$10 keV) & 1.7 & 1.7 
                                              & 1.7 \\
\tableline

\multicolumn{4}{l}{NOTE.--- Errors are 90\% confidence level for a single
parameter.}\\
\multicolumn{4}{l}{Best-fit parameters are not corrected for the red shift 
(z = 0.04732)}\\
\multicolumn{4}{l}{$^a$ model 1 = power-law $\times$ Galactic absorption.}\\
\multicolumn{4}{l}{$^b$ model 2 = power-law $\times$ Galactic absorption, 
where N$_H$ is fixed as the Galactic value.}\\
\multicolumn{4}{l}{$^c$ model 3 = (power-law + Gaussian) $\times$ Galactic
absorption.}\\
\multicolumn{4}{l}{$^d$ Iron line center energy.}\\
\multicolumn{4}{l}{$^e$ Intrinsic line width is fixed.}\\
\multicolumn{4}{l}{$^f$ Calculated iron line flux.}\\
\multicolumn{4}{l}{$^g$ Equivalent width of the iron line.}\\
\multicolumn{4}{l}{$^h$ Calculated mean continuum flux in units of
10$^{-11}$ ergs cm$^{-2}$ s$^{-1}$.}\\

\end{tabular}
\end{center}
\end{table}

%-------------------------------------------------------------------
\clearpage

\begin{table}
\begin{center}
\caption{BEST-FITTING PARAMETERS FOR THE ASCA SPECTRUM}
\vspace*{0.7cm}

\begin{tabular}{lccc} \tableline \tableline
PARAMETERS &   & BEST-FIT VALUES & \\ \cline{2-4}
           &  GAUSSIAN LINE &  & DISK LINE\\ 
\tableline

$\Gamma$   & 1.73$^{+0.04}_{-0.03}$ & & 1.73$^{+0.03}_{-0.03}$ \\
N$_H^a$ (10$^{20}$ cm$^{-2}$)   & 2.3 & & 2.3 \\
Normalization (10$^{-3}$) & 9.0$^{+0.3}_{-0.2}$ & & 9.0$^{+0.2}_{-0.2}$ \\
E$_K$ (keV)  & 6.3$^{+0.7}_{-0.4}$ & & 6.1$^{+0.5}_{-0.4}$ \\
$\sigma$     & 0.9$^{+0.6}_{-0.3}$  & & \nodata \\
R$_{in}$ (GM/c$^2$)  & \nodata  & & $<$ 40 \\
R$_{out}$ (GM/c$^2$)$^b$ & \nodata  & & 1000 \\
Inclination (deg)$^b$  & \nodata  & & 90 \\
I$_{Fe}$ ($10^{-4}$ photons cm$^{-2}$ s$^{-1}$) & 1.8$^{+1.2}_{-0.6}$ & &
                                                  1.8$^{+0.8}_{-0.7}$\\ 
EW (keV)     & 0.5$^{+0.3}_{-0.2}$ & & 0.5$^{+0.2}_{-0.2}$\\
$\chi^2/\nu$   & 20.4/34 & & 21.1/34 \\ 
Flux$^c$ (2.0$-$10 keV) & 3.5 & & 3.5\\
\tableline

\multicolumn{4}{l}{NOTE.--- Errors are 90\% confidence level for a single
parameter. The red-shift effect is not corrected.}\\
\multicolumn{4}{l}{$^a$ Fixed as the value of the combined spectra 
(Table 2).} \\
\multicolumn{4}{l}{$^b$ Fixed in the spectral fitting.}\\
\multicolumn{4}{l}{$^c$ Calculated continuum flux in units of 
10$^{-11}$ ergs cm$^{-2}$ s$^{-1}$.}\\

\end{tabular}
\end{center}
\end{table}

%-------------------------------------------------------------------
\clearpage

\begin{table}
\begin{center}
\caption{BEST-FITTING RESULTS FOR THE ROSAT SPECTRA} 
\vspace*{0.7cm}

\begin{tabular}{lcccc} \tableline \tableline
PARAMETERS & 1991 November 21 & 1993 May 21-23 & 1993 May 24-25 &
1993 December 1\\ 
\tableline

$\Gamma$ & 2.10$^{+0.09}_{-0.09}$ & 2.1$^{+0.1}_{-0.1}$ &
           2.1$^{+0.2}_{-0.2}$    & 2.1$^{+0.4}_{-0.4}$ \\
N$_H$ (10$^{20}$ cm$^{-2}$) & 3.0$^{+0.3}_{-0.3}$ & 3.5$^{+0.5}_{-0.4}$ 
                            & 3.6$^{+0.7}_{-0.6}$ & 3.2$^{+1.8}_{-1.4}$
\\
Normalization (10$^{-3}$) & 14.1$^{+0.4}_{-0.3}$ & 3.4$^{+0.1}_{-0.1}$ 
                          & 3.8$^{+0.2}_{-0.2}$  & 2.5$^{+0.3}_{-0.3}$
\\
$\chi^2/\nu$ & 14.2/21 & 6.7/21 & 14.9/21 & 5.9/21 \\
Flux$^a$ (0.1$-$2.0 keV) & 39.3 & 9.2 & 10.1 & 7.0\\
\tableline

\multicolumn{5}{l}{NOTE.--- Errors are 90\% confidence level for a single
parameter.}\\
\multicolumn{5}{l}{$^a$ Calculated continuum flux in units of
10$^{-12}$ ergs cm$^{-2}$ s$^{-1}$.}\\

\end{tabular}
\end{center}
\end{table}

%-------------------------------------------------------------------
\clearpage
\noindent
\begin{center}
{\bf FIGURE CAPTIONS}
\end{center}

\figcaption[]{X-ray light curves of MCG$-$2-58-22 obtained from the ROSAT/PSPC
(left panels; 0.1$-$2.0 keV) and the ASCA/GIS observations (right panels;
0.8$-$10 keV\@). Data from GIS2 and GIS3 are averaged in the plots.
Light curves are calculated in 300 s bin for all the data sets, and no 
background was subtracted.}

\figcaption[]{Part of the ROSAT light curve of MCG$-$2-58-22 during the
1991 November 21 observation, when the source showed a very rapid time
variation.  Data were binned in 300 s intervals.}

\figcaption[]{Long-term X-ray light curve of MCG$-$2-58-22 obtained in a 
2$-$10 keV band. Filled and open circles in the figure represent the flux
data calculated in the current study and those adopted from the literature,
respectively. The dashed lines are included for the convenience of the 
interpretation.}

\figcaption[]{Average energy spectra of MCG$-$2-58-22 observed in 
1993 May 25--26. In this figure, each spectrum was rebinned appropriately 
to give better statistics per bin.   Then the model function, 
(power-law + Gaussian line) $\times$ low-energy absorption, was fitted to the 
PSPC, SIS and GIS spectra simultaneously. The solid lines represent the 
best-fit model function.}

\figcaption[]{Spectral ratio of the combined ROSAT/ASCA spectra to the 
continuum models.  The upper and lower panels take the model 1 and 
the model 2 (see text) as the reference, respectively.
Both panels show a weak and broad residual structure around 6$-$7 keV,
implying the presence of an iron line.  In addition, the lower panel 
clearly shows soft X-ray excess below about 0.5 keV to the model.}

\figcaption[]{Residuals of the ASCA spectra of 1997 June to the best-fit model
of the combined spectra. This figure displays a considerable discrepancy
between the SIS spectrum and the model below 1.5 keV, which
increases with decreasing energy.   The deviations are ascribed to the
reduction of quantum efficiency in SIS by radiation damage.}

\figcaption[]{Spectral ratio between the ASCA/SIS spectrum and the best-fit 
model continuum of the combined spectra (model 3 in Table 2).
This figure clearly shows the presence of a broad iron line.}

\figcaption[]{Spectral ratio of the ROSAT spectra to the best-fit model 
of the combined spectra (model 3 in Table 2).}

\figcaption[]{Confidence contours (68\%, 90\%, and 99\% levels) for 
power-law photon index vs.\ hydrogen equivalent column density for 
a fit to the ROSAT PSPC spectra.
Solid lines and short-dashed and long-dashed lines are for the spectra of
1991 November 21, 1993 May 21--25, and 1993 May 24--26, respectively.}

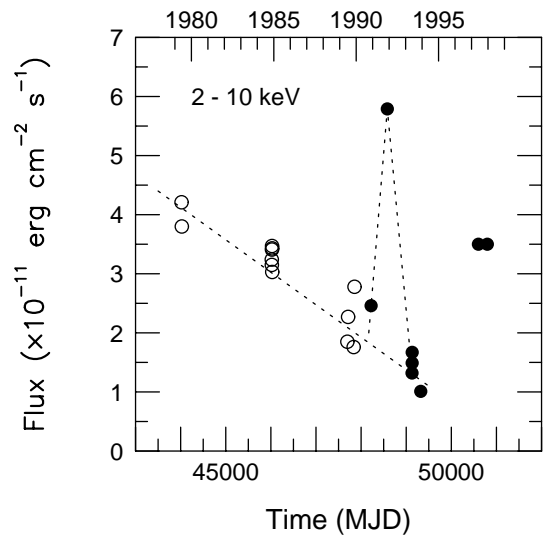
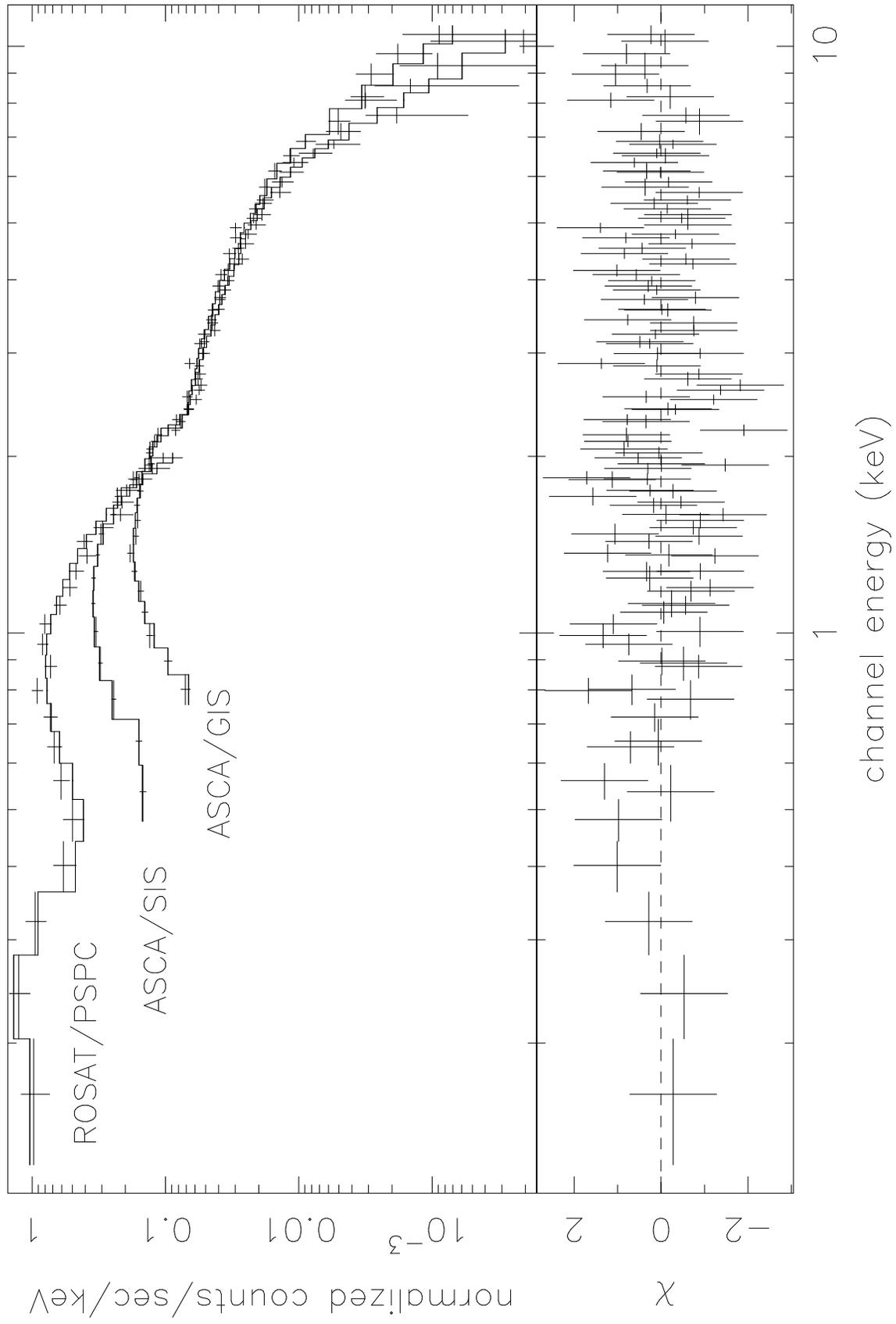
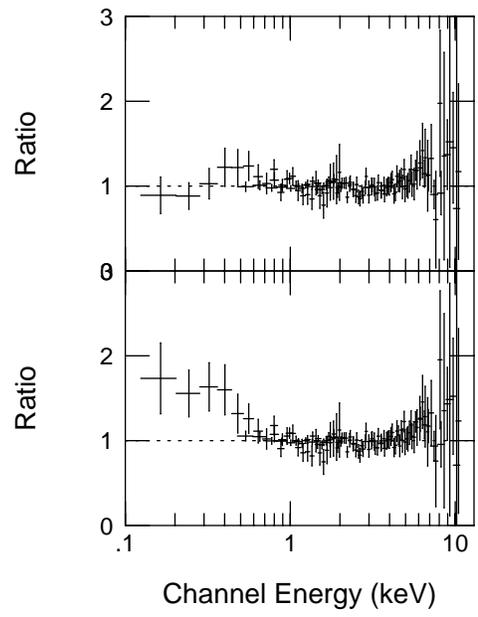
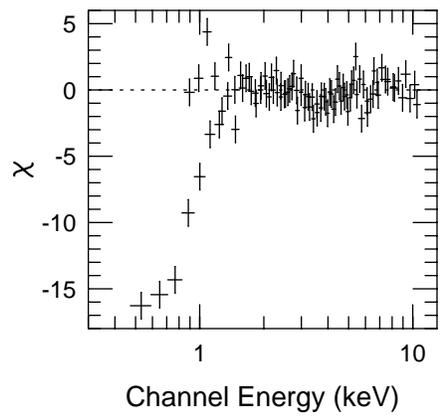
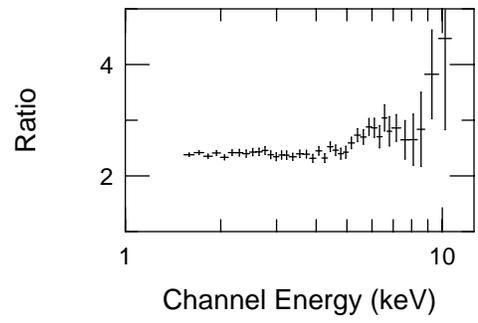
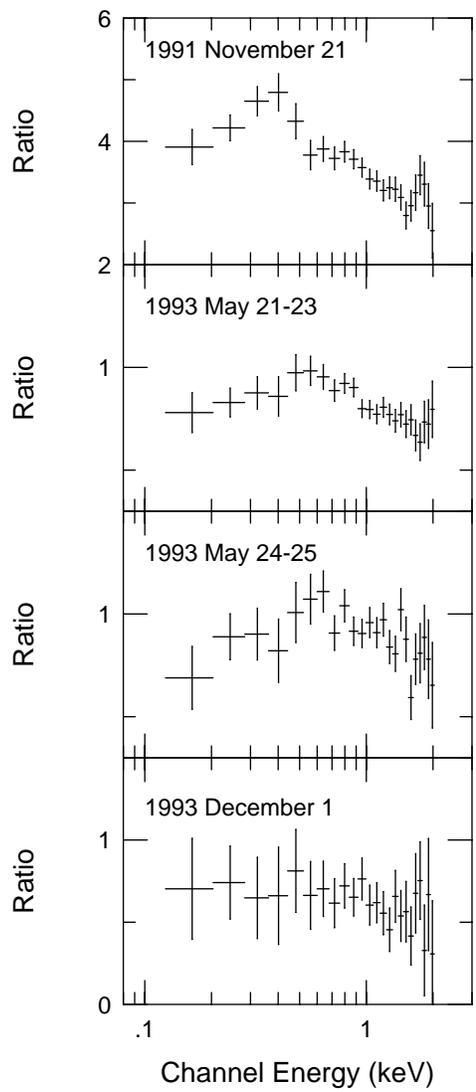
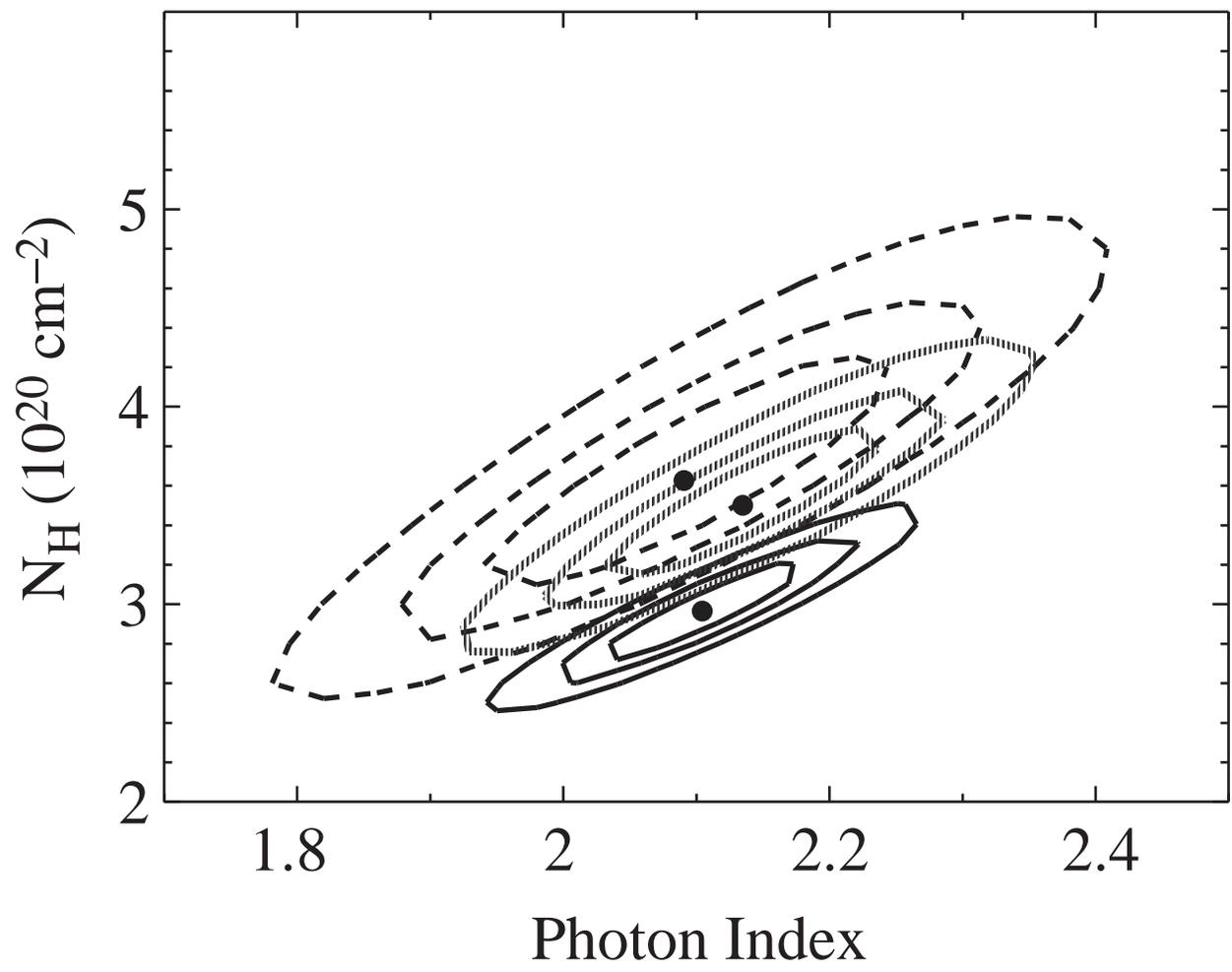
\end{document}